\documentclass{aastex63}

\usepackage{color}
\usepackage{wasysym}

\shorttitle{The mASC method for Terrestrial Exoplanets}
\shortauthors{Samara, Patsourakos $\&$ Georgoulis}

\begin{document}

\title{A Readily Implemented Atmosphere Sustainability Constraint for Terrestrial Exoplanets Orbiting Magnetically Active Stars}

\correspondingauthor{Evangelia Samara}
\email{evangelia.samara@kuleuven.be}

\author{Evangelia Samara}
\affiliation{Royal Observatory of Belgium, 1180 Brussels, Belgium}
\affiliation{Centre for Mathematical Plasma Astrophysics, KU Leuven, 3001 Leuven, Belgium}

\author{Spiros Patsourakos}
\affiliation{Department of Physics, University of Ioannina, 45110 Ioannina, Greece}

\author{Manolis K. Georgoulis}
\affiliation{Research Center of Astronomy and Applied Mathematics (RCAAM) of the Academy of Athens, 11527 Athens, Greece}

\begin{abstract}

\noindent With more than 4,300 confirmed exoplanets and counting, the next milestone in exoplanet research is to determine which of these newly found worlds could harbor life. Coronal Mass Ejections (CMEs), spawn by magnetically active, superflare-triggering dwarf stars, pose a direct threat to the habitability of terrestrial exoplanets as they can  deprive them from their atmospheres. Here we develop a readily implementable atmosphere sustainability constraint for terrestrial exoplanets orbiting active dwarfs, relying on the magnetospheric compression caused by CME impacts. Our constraint focuses on a systems understanding of CMEs in our own heliosphere that, applying to a given exoplanet, requires as key input the observed bolometric energy of flares emitted by its host star. Application of our constraint to six famous exoplanets, (Kepler-438b, Proxima-Centauri b, and Trappist-1d, -1e, -1f and -1g), within or in the immediate proximity of their stellar host's habitable zones, showed that only for Kepler-438b might atmospheric sustainability against stellar CMEs be likely. This seems to align with some recent studies that, however, may require far more demanding computational resources and observational inputs. Our physically intuitive constraint can be readily and \textit{en masse} applied, as is or generalized, to large-scale exoplanet surveys to detect planets that could be sieved for atmospheres and, perhaps, possible biosignatures at higher priority by current and future instrumentation. 

\end{abstract}


\section{Introduction} 
\label{sec:intro}

Planets beyond our solar system have become an object of fascination in recent decades, with nearly regular references in headlines and the popular media. Only recently have observational capabilities evolved to the point where potential terrestrial planets are detected around M-type dwarf stars. However, young M-type dwarfs are known to be magnetically active, often more than our middle-aged Sun. Superflares in them is a common occurrence \citep[e.g.,][]{maehara2012,armstrong2016} that should be resulting in fast and massive coronal mass ejections \citep[CMEs; see, e.g.,][and several others]{khodachenko_etal07,lammer2007,vidotto_etal11}. CMEs are gigantic, eruptive expulsions of magnetized plasma and helical magnetic fields from the solar and stellar coronae at speeds that may well surpass local Alfv\'{e}nic and sound speeds, severely but temporarily disrupting stellar winds and generating shocks around their bodily ejecta. 

Contrary to solar flares that are known since the 19th century \citep{carrington_1859}, CMEs were only observed well into the space age \citep{howard_06} due to their much fainter magnitude compared to the bright solar photospheric disk. Stellar CME detections are notoriously ambiguous, although recent efforts offer reasonable hope \citep{argiroffi_etal19}. However, in strongly magnetized stellar coronae CMEs are inevitable. In case of intense stellar magnetic activity and the existence of an atmosphere that shields a planet, extreme pressure effects by CMEs owning to stellar mega-eruptions can, under certain circumstances, cause intense atmospheric depletion via ionization-triggered erosion \citep[e.g.,][]{zendejas_etal2010}. In the solar system, results from NASA's Mars Atmosphere and Volatile Evolution (MAVEN) mission seem to establish that the sustained eroding effect of solar interplanetary CMEs (ICMEs) may be responsible for the thin Martian atmosphere \citep{jakosky2015} after the planet's magnetic field weakened.

Our objective here is the introduction of a practical and reproducible (magnetic) atmosphere sustainability constraint (mASC), reflected on a positive, rational number $\mathcal{R}$ and relying on CME and planetary magnetic pressure effects. $\mathcal{R}$ is a dimensionless ratio that, in tandem with the considered habitability zone \citep[HZ; e.g.,][]{kopparapu2013}, can provide an understanding of which terrestrial exoplanets warrant further screening for the existence of a possible atmosphere ($\mathcal{R} <1$), or otherwise ($\mathcal{R}>1$). Our mASC becomes fully constrained in case of tidally locked exoplanets \cite[e.g.,][]{grassmeier_etal04} by means of an estimated planetary magnetic field, while if no tidal locking is assumed the planetary magnetic field can be replaced by a known benchmark field, such as Earth's or other. Only magnetic pressure effects are taken into account in this initial study, but our mASC $\mathcal{R}$ can be generalized at will. Given that pressure effects are extensive and additive, however, adding more terms (i.e., kinetic, thermal) to the CME pressure will only increase the adversity of possible atmospheric erosion effects for a studied exoplanet \citep[e.g.,][]{Ngwira_etal14}. 

\section{(Magnetic) Atmosphere Sustainability Constraint (mASC) }
\label{sec:masc}

The magnetic activity of the exoplanets' host stars reflects on several observational facts, including the bolometric energy of their flares. From it, and assuming a Sun-as-a-Star analogue further reflected on the solar magnetic energy -- helicity (EH) diagram \citep{tziotziou_etal12}, we estimate the magnetic helicity of stellar CMEs and a corresponding stellar CME magnetic field based on the fundamental principle of magnetic helicity conservation in high magnetic Reynolds number plasmas \citep{patsourakos_etal16,patgeo17}. As explained in Appendix \ref{appendix:SunToStarAnalogue}, CMEs are inevitable in strongly magnetized stellar coronae as they relieve stars from excess helicity that cannot be shed otherwise. 

The near-star CME magnetic field is propagated self-similarly in the astrosphere until it reaches exoplanet orbits \citep{patgeo17}. The mASC introduced in this study achieves a precise, quantitative assessment of whether the magnetic pressure of stellar ICMEs alone can be balanced by the estimated (tidally locked) or guessed (in the general case) magnetic pressure of \emph{observed} terrestrial exoplanets at a magnetopause distance large enough to avert erosion effects of a possible atmosphere. 

There are two conceptual pillars of the mASC introduced here: first, it relies on observed stellar flare energies but does not perform a case-by-case stellar eruption analysis. In other words, it does not look at the particular eruption but points to the collective effect of numerous similar eruptions over the $\sim$1 Gyear, or significant fractions thereof, of the young star's activity. In this sense, the suitable orientation required for an enhanced ICME planeto-effectiveness, namely, the ICMEs' ability to reconnect with the planetary magnetic field causing magnetic storms, is ignored: it is implicitly assumed that numerous such ICMEs will have the correct orientation. Second, our mASC ratio relies explicitly on a worst-case scenario
(i.e., largest possible ICME magnetic field strength)
magnetic pressure for stellar CMEs ($\sim B_{worst}^2$) \emph{and} a best-case scenario planetary magnetic field 
(i.e., largest possible planetary magnetic field)
generated in the planet's core due to internal dynamo action ($\sim B_{best}^2$). Then, our mASC becomes the ratio of planetary magnetic intensities relating to these pressure terms: the minimum planetary magnetic field (equal to $B_{worst}$) able to balance the worst-case CME magnetic field and the best-case planetary magnetic field $B_{best}$ as per the modeled planetary characteristics, i.e.,
\begin{equation}
    \mathcal{R} = {{B_{worst}} \over {B_{best}}}
    \label{eq:masc}
\end{equation}

The two planetary magnetic field strengths are taken at a critical magnetopause distance from the studied planet in terms of atmospheric erosion effects (Sections \ref{sec:bworst}, \ref{sec:bbest}). It is then understood that if $\mathcal{R}>1$, the planet's magnetosphere will be compressed beyond the critical threshold, presumably leading to atmospheric erosion (after processes such as thermal, nonthermal or hydrodynamic escape, catastrophic erosion and others, take place; see, for example, \citet{Melosh&Vickery1989,Lundin2004,Barabash2007} for more details). 
Assuming that, statistically, the presumed ICME is not a unique occurrence, the planet may undergo this atmospheric stripping for hundreds of millions of years due to its star's magnetic activity. Magnetic helicity conservation, on the other hand, dictates that at least some (or even most, or all) magnetic eruptions in the star will unleash powerful CMEs to shed away the excess helicity generated in the star's magnetized atmosphere that is otherwise conserved and remains accumulated in the star's corona. In such situations, one casts doubt on the viability of an atmosphere in the otherwise terrestrial planet, even if the planet is seated well into the HZ of its astrosphere. The opposite is the case for $\mathcal{R}<1$. 

\subsection{The worst-case CME-equipartition magnetic field}
\label{sec:bworst}

A critical magnetopause distance of $r_{mp}=2\;R_{p}$ (i.e., two planetary radii, or one radius away from the surface of the planet, see \citet{khodachenko_etal07,lammer2007}) was adopted as the minimum planetocentric distance in which atmospheric ionization and erosion can still be averted during extreme magnetospheric compression. Then, the \emph{equipartition} planetary magnetic field $B_{eq}$ (to be viewed as $B_{worst}$) that balances the ICME magnetic pressure at $r_{mp}=2\;R_{p}$ can be estimated as (see Appendix \ref{appendix:ICMEequipartition} for a complete description)

\begin{equation}
B_{eq} = 8\;B_{icme}\;\;,
\label{eq:eqf}
\end{equation}

\noindent where $B_{icme}$ is the worst-case ICME axial magnetic field at $r_{mp}=2R_p$.

To infer $B_{icme}$ at any given astrocentric distance $r_{icme}$, we first need to constrain the axial magnetic field $B_0$ of the CME at a near-star distance $r_0$ (see Appendix \ref{appendix:SunToStarAnalogue} for a derivation). $B_0$ is constrained by observational facts and more specifically by assuming a given flux rope model and a corresponding magnetic helicity formula depending on the radius $R$ and length $L$ of the flux rope, that can then be solved for $B_0$ \citep{patsourakos_etal16}. \citet{patgeo17} tested several linear (LFF), nonlinear (NLFF) and non-force-free flux rope models and determined that the worst-case scenario $B_0$ for near-Sun CME flux ropes was obtained by the LFF Lundquist flux-rope model (values estimated by other models range between 2-80 $\%$ of the Lundquist values) that gives a magnetic helicity of the form

\begin{equation}
H_m = {{4 \pi B_0^2 L} \over {\alpha}} \int _0^R J_1^2(\alpha r)rdr\;\;,
\label{eq:hm0}
\end{equation}

\noindent where $\alpha$ is the constant force-free parameter and $J_1()$ is the Bessel function of the first kind. Parameter $\alpha$ is inferred by the additional constraint $\alpha R \simeq 2.405$, imposed by the first zero of the Bessel function of the zero-th kind, $J_0()$, in the Lundquist model. 
The flux rope radius $R$ corresponds to the CME front
that is assumed to have a circular cross-section with maximum area. 

We use the Lundquist flux-rope model throughout this analysis, as this is a standard ICME model for the inner heliosphere. It gives the strongest near-Sun CME axial magnetic field $B_0$, provided that the twist is not excessive (see \citet{patgeo17}). Adopting the fundamental helicity conservation principle for high magnetic Reynolds number plasmas \citep[e.g.,][]{berger84} implies a fixed $H_m$ and dictates that as the CME expands, $B_0$ decreases self-similarly as a function of $1/r^2$, with $r$ being the heliocentric distance. We maintain this quadratic scaling for distances relatively close to the Sun (e.g., up to the Alfv\'enic surface where the solar wind speed matches the local Alfv\'en speed at $\sim 10\;R_{\odot}$). Beyond that surface this analysis continues to adhere to helicity conservation but adopts a power-law radial fall-off index
$a_B = 1.6$ for the propagation of Lundquist-flux-rope solar CMEs within the astrospheres (see Appendix \ref{appendix:SunToStarAnalogue} for a derivation of the exponent). As a result, $B_{icme}$ at a given astrocentric distance $r_{icme}$ is given by 

\begin{equation}
B_{icme} = B_0 ({{r_0} \over {r_{icme}}})^{1.6}    
\label{eq:prop}    
\end{equation}

\noindent where $r_0$ is the near-star distance up to which the CME axial magnetic field scales as $(1/r^2)$ and $B_0$ is this magnetic field at that distance.

\subsection{The best-case planetary magnetic field} 
\label{sec:bbest}

The 'defense' line in the ICME pressure effects for any given planet is being held primarily by the planet's magnetic pressure. The planet's dipole magnetic moment $\mathcal{M}$ gives rise to a planetary magnetic field 
 
\begin{equation}
B_p = {{\mathcal{M}} \over {r_{mp}^3}}
\label{eq:bp}
\end{equation}

\noindent for the dayside magnetopause occurring at a planetocentric distance $r_{mp}$. To identify the best-case scenario, we examined several prominent models for the magnetic moment $\mathcal{M}$ \citep{busse76,stevenson_etal83,mizutani_etal92,sano93} --see also \citet{christensen10} for a review-- to determine which would provide the strongest $\mathcal{M}$, focusing particularly on the tidally locked regime. We concluded that an upper-case $\mathcal{M}$ is provided by \citet{stevenson_etal83} and a model variant of \citet{mizutani_etal92}, namely

\begin{equation}
{\mathcal{M}} = \mathcal{M}_{Stev} \simeq A\;\rho _c^{1/2} \omega ^{1/2} R_c^3\;\sigma ^{-1/2}.
\label{eq:mmom}
\end{equation}

\noindent The other models provided values ranging between 18-62 $\%$ of the Stevenson value. In Equation (\ref{eq:mmom}), $A \simeq 3.45 \times 10^5$ A$\cdot$s$\cdot$kg$^{-0.5}$ is the proportionality constant, $\omega$ corresponds to the planet's angular rotation and $\rho _c$, $R_c$ and $\sigma$ correspond to the planetary core's mean density, radius and electrical conductivity, respectively (see Appendix \ref{appendix:PlanetaryBinTidallyLockedExoplanets} for more details 
on the calculations and
assumptions made). $B_p$ for $\mathcal{M}$ = $\mathcal{M}_{Stev}$ 
($B_{Stev}$, hereafter) is to be viewed as $B_{best}$. \

Summarizing, our mASC ratio of Equation (\ref{eq:masc}) translates to $\mathcal{R}=(B_{eq}/B_{Stev})$. $B_{eq}$ can be estimated in case of known bolometric stellar flare energies, while $B_{Stev}$ is fully constrained for tidally locked exoplanets.

\section{Application of the mASC method}

\begin{figure}[ht]
    \centering
    \includegraphics[width=0.8\textwidth]{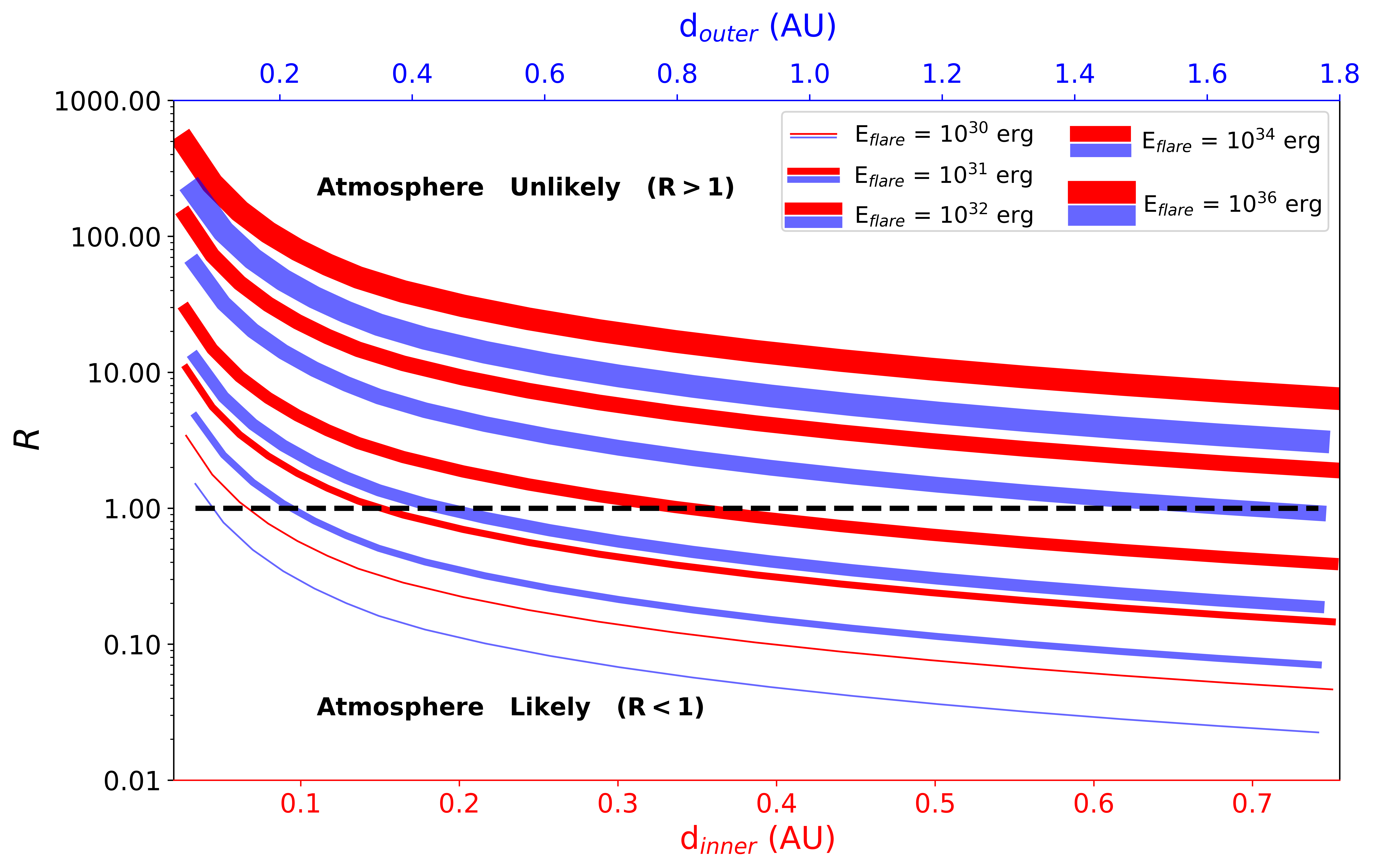}
    \caption{The mASC ratio $\mathcal{R}=(B_{eq}/B_{Stev})$ for Earth twins, but with different equatorial magnetic fields, lying on the inner (red) and outer (blue) HZ boundary inferred by \citet{kopparapu2013} for different stellar flare energies, each represented by a different curve thickness. The astrocentric distance in the abscissas implicitly includes the stellar mass shown in Figure \ref{fig:chz}. The limit $\mathcal{R}=1$ (dotted lines) separates an apparent non-viability of an atmosphere  
    ($\mathcal{R} > 1$) from an apparently likely atmosphere ($\mathcal{R} < 1$) for both cases.}
    \label{fig:inouthz}
\end{figure}

\begin{figure}[ht]
    \centering
    \includegraphics[width=0.7\textwidth]{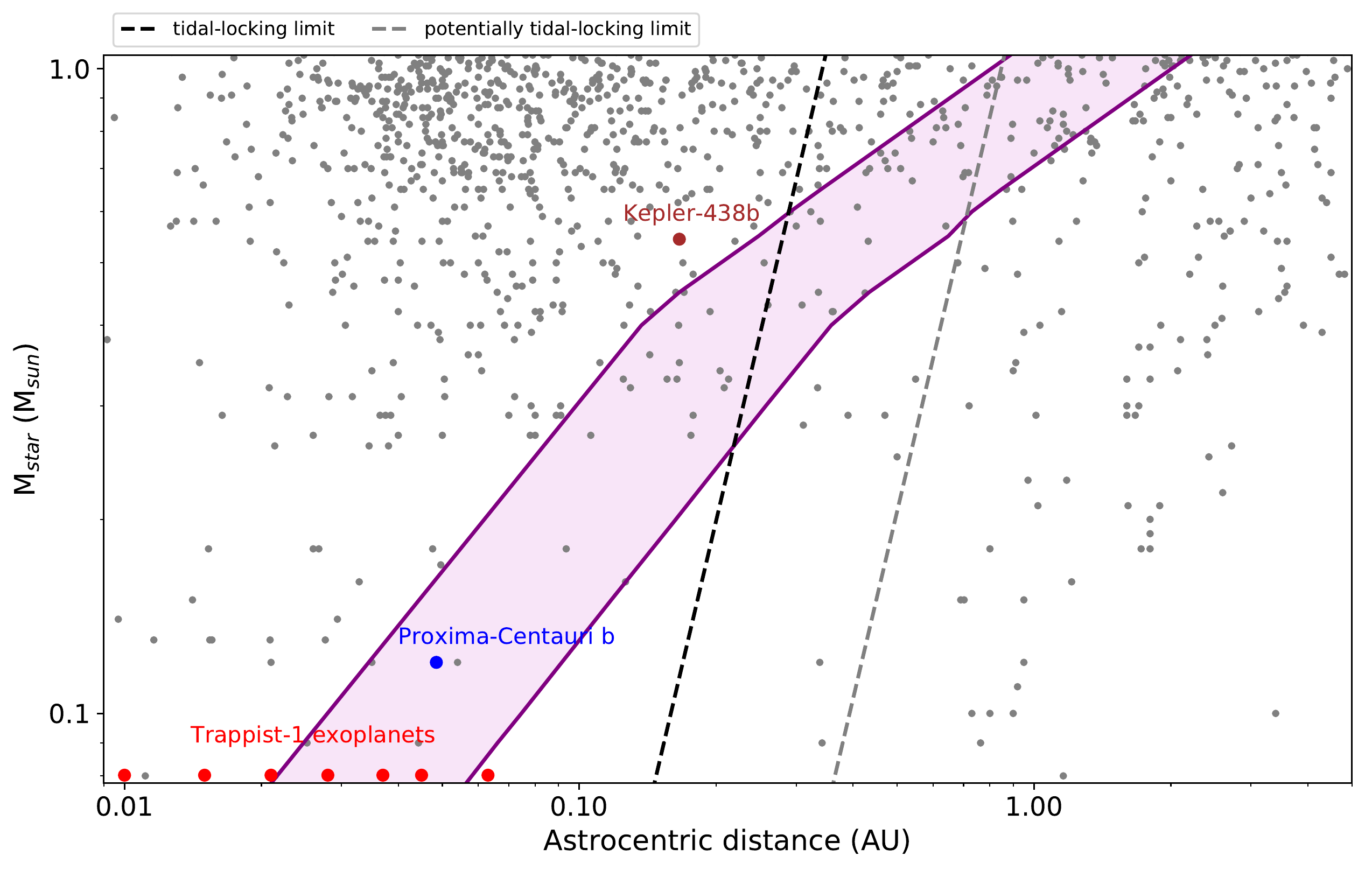}
    \caption{An ensemble of 1,771 confirmed exoplanets by the NASA Exoplanet Archive plotted on a diagram of stellar mass vs. astrocentric distance. We restrict the analysis to stars less or equally massive to the Sun while exoplanets shown have a confirmed orbital semi-major axis. Inner and outer HZ limits are indicated by solid lines enclosing the pink-shaded HZ area, while the black and gray dashed lines indicate internal and potential external tidal-locking limits, respectively. The locations of nine exoplanets, six of which are studied here due to their proximity to, or presence within, the HZ, are also highlighted.}
    \label{fig:chz}
\end{figure}

Assuming tidal locking to fully determine $B_{Stev}$, Figure \ref{fig:inouthz} provides the nominal values of $\mathcal{R}$ for different stellar flare energies and for an Earth-like exoplanet lying precisely at the inner (red) and outer (blue) HZ boundary of \citet{kopparapu2013}. This plot represents a different conception of Figure \ref{fig:chz} that shows a number of exoplanets provided by the NASA Exoplanet Archive lying within and without the tidally locked regime and within and without the inner and outer HZ limits. Nevertheless, Figure \ref{fig:inouthz} now includes $\mathcal{R}$ as a function of stellar flare energies while stellar masses are implicitly included in the astrocentric distances $d_{inner}$ and $d_{outer}$. It comes as no surprise that higher flare energies, statistically resulting in more helical CMEs and stronger axial magnetic fields, require the planet to be orbiting its host star at a larger orbital distance to achieve $\mathcal{R} < 1$. For flare energies higher than $10^{33}$ erg it appears that planets located precisely on the inner HZ boundary may be incapable to sustain an atmosphere while this is the case for energies higher than $10^{34}$ erg for planets located on the outer HZ boundary. 

\begin{figure}[h!]
    \centering
    \includegraphics[width=0.95 \textwidth]{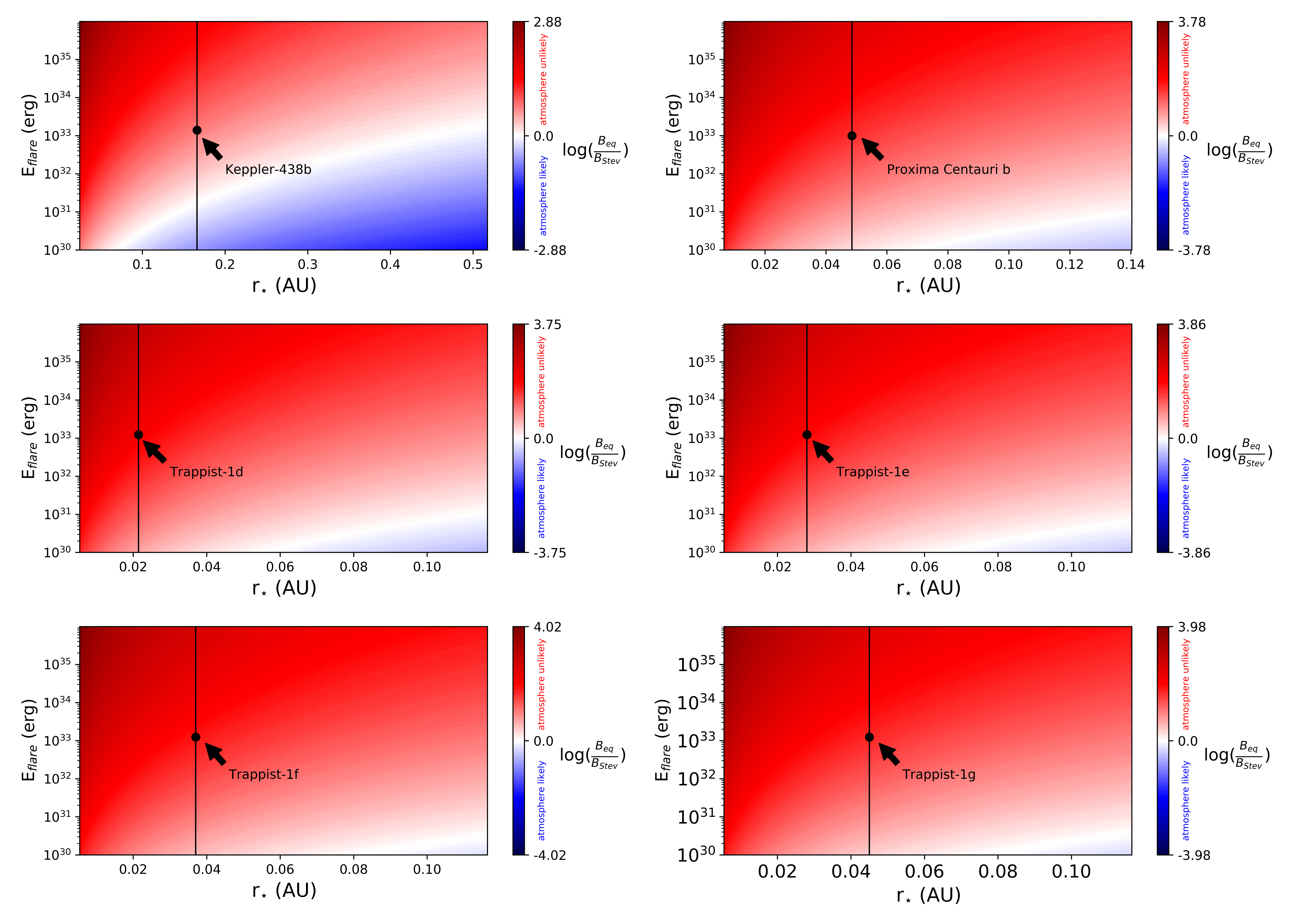}
    \caption{Testing the viability of a potential atmosphere from the value of the mASC $\mathcal{R}$ for six confirmed exoplanets, namely (clockwise from top left) K438b, PCb, Tp1e, Tp1g, Tp1f and Tp1d. Ordinates correspond to stellar flare energies and abscissas to astrocentric distances. The color scales correspond to $\log({\mathcal{R}})$. Each planet is represented by a point corresponding to coordinates set by the maximum observed flare energies from its host star and confirmed orbital distances, highlighted by vertical  lines.}
\label{fig:cbc}
\end{figure}

Importantly, our mASC method can also be applied to case studies of terrestrial exoplanets, provided that flares from their host stars are observed. If these planets are --or are assumed to be-- tidally locked, then $\mathcal{R}$ becomes fully constrained. In Figure \ref{fig:cbc}, we examine six popular cases of terrestrial exoplanets, all within the tidal-locking zone and either within, or close to, the respective HZ of their host stars. These cases are Kepler-438b (K438b), Proxima-Centauri b (PCb), and four Trappist-1 (Tp1) exoplanets, namely Tp1d, Tp1e, Tp1f and Tp1g. Actual mASC $\mathcal{R}$-values and applicable uncertainties (see Appendix \ref{appendix:SensitivityAnalysis}) are provided
in Table \ref{tab:pspecs}. Figure \ref{fig:cbc} offers a graphical representation of these results where observed flare energies from host stars have been taken from \citet{armstrong2016}, \citet{Howard2018} and \citet{vida2017} for Keppler-438, Proxima Centauri and Trappist-1, respectively.

The uncertainty analysis mentioned above and described in Appendix \ref{appendix:SensitivityAnalysis} aims
to determine under which circumstances could a conclusion of $\mathcal{R}<1$ or $\mathcal{R}>1$ be reversed due to applicable uncertainties. In particular, we define an \emph{equipartition} scaling index $a_{B_{(\mathcal{R}=1})}$ for which $\mathcal{R}=1$, along with its uncertainty $\delta a_B$. If (i) $|a_B - a_{B_{(\mathcal{R}=1})}| > \delta a_B$ for the nominal $a_B = 1.6$ of \citet{patgeo16,patgeo17}
or (ii) $a_{B_{(\mathcal{R}=1})}$ is steeper than 2 beyond applicable $\delta a_B$, with 2 being the 'vacuum' radial fall-off index near the star and $a_B <2$ reflecting an astrosphere with its own MHD environment, then our conclusion of $\mathcal{R}<1$ or $\mathcal{R}>1$ is unlikely to change. If uncertainties are large enough to preclude a safe conclusion, then our main result on an exoplanet is likely to change because of these uncertainties. Analogous uncertainties may be sought in the case of additional pressure effects included in $\mathcal{R}$. 

Table \ref{tab:pspecs} shows that for all six presumed tidally locked exoplanets the 1:1 spin-orbit resonance results in planetary rotations that are slow enough to allow $\mathcal{R}>1$, rendering a sustainable atmosphere unlikely. In all cases, $a_{B_{(\mathcal{R}=1)}}$ is well above the 'vacuum' value of 2, with PCb and the four Tp-1 exoplanets showing $a_{B_{(\mathcal{R}=1)}} >2$ beyond applicable uncertainties. Hence, for five out of six potentially tidally locked exoplanets, our result for $\mathcal{R} >1$, meaning a non-sustainable atmosphere, seems robust. The result could be reversed for K438b as the difference between $a_{B_{(\mathcal{R}=1)}}$ and 1.6 is within applicable uncertainties, meaning that if $a_B$ systematically lies in the range $(1.6, 2.0)$ then the exoplanet might conceivably sustain an atmosphere.

\begin{table}[h!]
    \centering
    \begin{tabular}{c c c c c c c}
    Exoplanet & Abridged & $\mathcal{R}$ & Atmosphere &  
    $a_{B_{(\mathcal{R}=1)}}$ & $\delta a_B$ & Result\\
    & & & likely? & & & robust?\\
    \hline 
    Keppler-438b & K438b & 5.46 & No & 2.48 & 0.99 & No \\
    Proxima Centauri b & PCb & 42.84 & No & 3.48 & 0.96 & Yes \\
    Trappist-1d  & Tp1d & 95.35 & No & 4.95 & 1.40 & Yes\\
    Trappist-1e  & Tp1e & 77.87 & No & 4.24 & 1.15 & Yes \\
    Trappist-1f  & Tp1f & 70.61 & No & 3.80 & 0.99 & Yes\\
    Trappist-1g  & Tp1g & 48.53 & No & 3.43 & 0.90 & Yes\\
    \hline
    \end{tabular}
    \caption{Six case studies of the mASC $\mathcal{R}$-value as depicted in Figure \ref{fig:cbc}. The values of $\mathcal{R}$ correspond to our nominal radial power-law fall-off index  $a_B=1.6$. The index $a_{B_{(\mathcal{R}=1)}}$  corresponds to the equipartition $a_B$-value for which $\mathcal{R}=1$, while $\delta a_B$ corresponds to the uncertainty of $a_{B_{(\mathcal{R}=1)}}$. The last column assesses whether our result is robust as per the difference $a_{B_{(\mathcal{R}=1)}} \pm \delta a_B$ from the nominal $a_B$-value.}
    \label{tab:pspecs}
\end{table}

The above underline the potential value of the mASC $\mathcal{R}$, even in its simplest form involving only magnetic pressure effects: lying in the HZ of their host stars does not necessarily make exoplanets capable of sustaining an atmosphere, a favorable Earth Similarity Index (ESI; \citet{schulze_etal11}) notwithstanding. This would be the case for PCb and Tp1d,e,f, and g although virtually all lie in the HZ. Conversely, K438b is slightly beyond the inner HZ that might inhibit an atmosphere and possible liquid water on its surface but, at the same time, due to its larger orbital distance it might be relatively immune to at least plausible, as per flare observations, space weather from its host star.

\section{Conclusions}

This versatile and highly reproducible analysis shows that space weather cannot be left out of considerations for planetary habitability in stellar systems. It carries both value and promise: consider the European Space Agency CHEOPS (Characterizing Exoplanet Satellite) mission, for example \citep[ e.g.,][]{sulis_etal20}. The mission is designed to characterize only \emph{selected}, confirmed exoplanets, ranging from super-Earth to Neptune sizes, aiming toward studies that extend into their potential atmospheres. Although the mission is already in orbit, analyses such as this could help assess future observing priorities. The same applies to optimizing exoplanet selection for biosignature analysis in the framework of the upcoming James Webb mission.

In anticipation of these exciting observations that will give the ultimate
test to our method, we hereby supply a first round of tentative tests aiming towards its validation.
For example, it was recently shown that LHS 3844b \citep{Vanderspek2019, Kane2020}, a rocky exoplanet orbiting a M-dwarf star within its tidally locked zone, lacks an atmosphere. \citet{Kane2020} suggest that the mother star of LHS 3844b exhibited an active past, comparable to that of Proxima Centauri. By adopting, therefore, a maximum super-flare energy identical to the one of Proxima Centauri, we obtained R = 251.08 ($>$ 1, atmosphere unlikely) which agrees with the observations. Also, our result seems robust because $|a_B - a_{B_{(\mathcal{R}=1})}| > \delta a_B$.\footnote{For this specific exoplanet which orbits very close to its mother star, we maintain the quadratic scaling of $B_0$ with astrocentric distance ($~ 1/r^2$) for distances up to 7$R{_*}$ and not 10$R{_*}$. Beyond 7$R{_*}$, we adopt the same power-law radial fall-off index $a_B$  = 1.6 for the propagation of Lundquist-flux-rope solar CMEs within the astrospheres.} Our mASC could be indirectly validated as well, by checking if other studies -- focusing also on tidally locked, terrestrial worlds and having similar objectives -- converge on similar results. This said, results presented here are in qualitative agreement with MHD simulations of stellar winds, for example by \citet{garaffo_etal17}, that find extreme magnetospheric compressions below $\approx$ 2.5 planetary radii for Tp1d-g and a magnetopause distance of $[1.5,4.5]$ planetary radii for PCb. Recent extensive (and computationally expensive) MHD models of stellar CMEs \citep[e.g.,][]{lynch2019} along with semi-empirical approaches \citep[e.g.,][]{kay2016} have emerged, and both approaches take as input maps of the stellar surface magnetic field inferred by Zeeman-Doppler imaging reconstructions \citep[e.g.,][]{donati1997}. While detailed, these studies may be rather impractical for bulk application to a large number of exoplanets. On the other hand, our introduced mASC could provide guidance to large-scale MHD simulations of stellar CMEs, by efficiently scanning and bracketing the corresponding parameter space so that the simulations could be performed only to pertinent cases. 

Concluding, we reiterate that mASC $\mathcal{R}$ can be generalized at will with additional pressure terms, albeit mainly from the ICME side. In its most general form, $\mathcal{R} = P_{eq} / P_{planet}$, with worst- and best-case scenario pressure terms $P_{eq}$ and $P_{planet}$, respectively. We note, in particular, the study of \citet{moschou2019} where a (flare) energy vs. (CME) kinetic energy diagram is inferred from stellar observations and modeling. Such statistics could be integrated into the energy-helicity diagram of this study to revise the $P_{eq}$-term. The energy-helicity diagram used here is also an entirely solar one, so any possibilities to extend it to better reflect the magnetic activity of dwarf, planet-prolific stars are well warranted. Equally meaningful $P_{planet}$-terms could be introduced to provide a far more sophisticated, but still readily achieved, mASC ratio $\mathcal{R}$ for the screening of alien terrestrial worlds for potential atmospheres and, ultimately, life.

\acknowledgments
The authors would like to thank the anonymous referee for his comments and suggestions that improved the manuscript. This work was inspired by and originated during the M.Sc. Thesis of ES, implemented at the University of Athens and the Research Center for Astronomy and Applied Mathematics of the Academy of Athens, Greece. We thank both institutions for their support and encouragement. The authors would also like to extend their acknowledgements to Prof. Dr. Stefaan Poedts from the Centre of Mathematical Plasma-Astrophysics (CMPA) KU Leuven, for his support and great incitement on this work. 

\vspace{5mm}
\facilities{The properties of the exoplanets and host-stars used in this study are available at the NASA Exoplanet Archive (https://exoplanetarchive.ipac.caltech.edu/).}

\software{The code to implement the mASC for any terrestrial-like exoplanet lying within the tidally locked regime of its host star is available in the following repository: https://github.com/SamaraEvangelia/mASC-method}

\clearpage

\appendix

\section{Near- Sun/star CMEs and helio-/astro-spheric propagation}
\label{appendix:SunToStarAnalogue}

A physically meaningful expression of magnetic helicity in the Sun and magnetically active stars is the relative helicity, related to the absence of vacuum and hence the flow of electric currents in the solar and stellar coronae, along with the topological settings of solar (and stellar) magnetic fields that are only partially observed and detected, on and above the stars' surface. By construction, the relative helicity must be quantitatively connected to the excess or free magnetic energy that is also explicitly due to the presence of electric currents \citep[e.g.,][]{sakurai81}. An attempt to correlate the free magnetic energy with the relative magnetic helicity in solar active regions was taken by \citet{georgoulis_labonte07} for linear force-free (LFF) magnetic fields, and then by \citet{georgoulis_etal12} for nonlinear force-free (NLFF) ones.

The NLFF (magnetic) energy - (relative) helicity correlation resulted in the energy - helicity (EH) diagram of \citet{tziotziou_etal12}. There, $\sim$160 vector magnetograms of observed solar active regions were treated in the homogeneous way of \citet{georgoulis_etal12}, aiming toward a scaling between the NLFF free magnetic energy $E_c$ and relative helicity $H_m$. They found a robust scaling of the form (CGS units)

\begin{equation}
\log \mid H_m \mid \propto 53.4 - 0.0524 (\log E_c)^{0.653} 
\exp{ {{97.45} \over {\log E_c}}}\;\;. 
\label{eq:eh_1}
\end{equation}

\noindent Here $\mid H_m \mid$ refers to the magnitude of the relative helicity $H_m$, that can be right- (+) or left- (--) handed. A similar expression to Equation (\ref{eq:eh_1}) provides a simpler, power-law dependence between $\mid H_m \mid$ and $E_c$, of the form (CGS units)

\begin{equation}
\mid H_m \mid \propto 1.37 \times 10^{14} E_c^{0.897}. 
\label{eq:eh_2}
\end{equation}

The above EH scaling was shown to hold for typical active-region free energies in the range $E_c \sim (10^{30}, 10^{33})$ erg and respective relative helicity budgets $\mid H_m \mid \sim (10^{40}, 10^{44})$ Mx$^2$. The robustness of the EH diagram scaling was validated in multiple cases that involved not only active regions but also quiet-Sun regions and magnetohydrodynamic models \citep{tziotziou_etal13}. 

Combining the EH scaling with the conservation principle of the relative magnetic helicity, $E_c$ in Equations (\ref{eq:eh_1}) and (\ref{eq:eh_2}) is dissipated in every instability (solar flare, CME, etc.), while $H_m$ is only shed away from the Sun via CMEs. If an active region was to completely relax (i.e., return to the vacuum energy state) by a \textit{single} magnetic eruption, then for a given free magnetic energy $E_c$ it would expel helicity $H_m$. This is the core reasoning behind a worst-case-scenario solar eruption originating from a given solar source. Typically, up to $\sim$10\% of the total free energy and up to 30 -- 40\% of the total magnetic helicity of the source are dissipated and ejected, respectively, in a solar eruption \citep[see, e.g.,][]{nindos_etal03, moriatis_etal14}. 

The majority of CME ejecta are observed to be in the form of a magnetic flux rope \citep[e.g.,][]{vourlidas_etal13, vourlidas_etal17}, with this geometry surviving the CMEs' inner heliospheric propagation all the way to the Sun-Earth libration point L1 and probably beyond \citep[e.g.,][]{zurbuchen_richardson06, nieves_etal18}. Based on the flux rope CME geometry, tools such as the Graduated Cylindrical Shell (GCS) model of \citet{thernisien_etal09, thernisien11}, processed observations by the STEREO/SECCHI coronagraph \citep{howard_etal08} to obtain the aspect ratio $k=R/L$ between the radius $R$ and along with the half-angular width $w$ of the CME flux rope at some distance (typically, 10 solar radii, $R_{\odot}$) away from the Sun. The length $L$ corresponds to the perimeter of the flux rope while the radius $R$ corresponds to the CME front, that is assumed to have a circular cross-section with maximum area.

The worst-case scenario $B_0$ for near-Sun CME flux ropes was provided by the LFF Lundquist flux-rope model which was used throughout this analysis and gives a magnetic helicity of the form 

\begin{equation}
H_m = {{4 \pi B_0^2 L} \over {\alpha}} \int _0^R J_1^2(\alpha r)rdr\;\;,
\label{eq:hm0}
\end{equation}

\noindent where $\alpha$ is the constant force-free parameter and $J_1()$ is the Bessel function of the first kind. Parameter $\alpha$ is inferred by the additional constraint $\alpha R \simeq 2.405$, imposed by the first zero of the Bessel function of the zero-th kind, $J_0()$, in the Lundquist model.

As already mentioned, for helicity conservation imposing a fixed $H_m$, the Lundquist model requires that as the CME expands, $B_0$ decreases self-similarly as a function of $1/r^2$ for distances close to the Sun. In this analysis however and for distances away from the Sun (see also \citet{patgeo16}), the self-similar expansion was not a priori assumed to be quadratic (i.e., $1/r^2$), but more generally, $(1/r^{a_B})$, with $a_B$ being the absolute value of the power-law radial fall-off index. This different index, still under helicity conservation that dictates respective power laws for the increase (expansion) of the CME flux rope $R$ and $L$, aimed to include all effects present during heliospheric propagation and the interaction of ICMEs with the ambient solar wind \citep[see, e.g.,][for an account of these effects]{machester2017}. Prominent among them is the CME flattening \citep[see, e.g.,][and references therein]{raghav_shaikh20} that tends to distort the CME geometry due to plasma draping or flux-pileup as the CME pushes through the heliospheric spiral. As a result, inner-heliospheric propagation implies that $B_{icme}$ at a given heliocentric distance $r_{icme}$ is given by 

\begin{equation}
B_{icme} = B_0 ({{r_0} \over {r_{icme}}})^{a_B}    
\label{eq:prop_appendix}    
\end{equation}
where $r_0$ is the (near-Sun) distance up to which the CME axial magnetic field scales as $(1/r^2)$ and $B_0$ is this magnetic field at that distance.

This analysis adopts $a_B = 1.6$ in Equation (\ref{eq:prop_appendix}) for the propagation of Lundquist-flux-rope stellar CMEs within their astrospheres (where $a_B$ can vary in the range [1.34, 2.16], see \citet{salman2020}). A Monte Carlo simulation for various stellar $(k,w)$-pairs is adopted, while the stellar CME $\mid H_m \mid$ is inferred by the bolometric, observed stellar flare energies via Equation (\ref{eq:eh_2}). A valid question is where in the near-star space are we to apply the model $(k,w)$, that were taken at $10\;R_{\odot}$: assuming a $10\;R_*$ astrocentric distance, where $R_*$ is the radius of the host star, we should apply a 'fudge-factor' correction to $B_0$ as follows: 

\begin{equation}
B_{0_{(10\;R_*)}} = B_{0_{(10\;R_{\odot})}} ({R_{\odot} \over {R_*}})^2\;\;.
\label{ff}
\end{equation}

\noindent This factor is adopted for the cases of this study's six exoplanets and their host stars. In the general case, or where no assessment is taken for the stellar radius, one may start the astrospheric propagation at a physical astrocentric distance of $10\;R_{\odot}$, independently from  the stellar radius (provided, of course, that this radius does not exceed 10 $R_{\odot}$).  

\begin{figure}[h!]
\centering
    \includegraphics[width=0.7\textwidth]{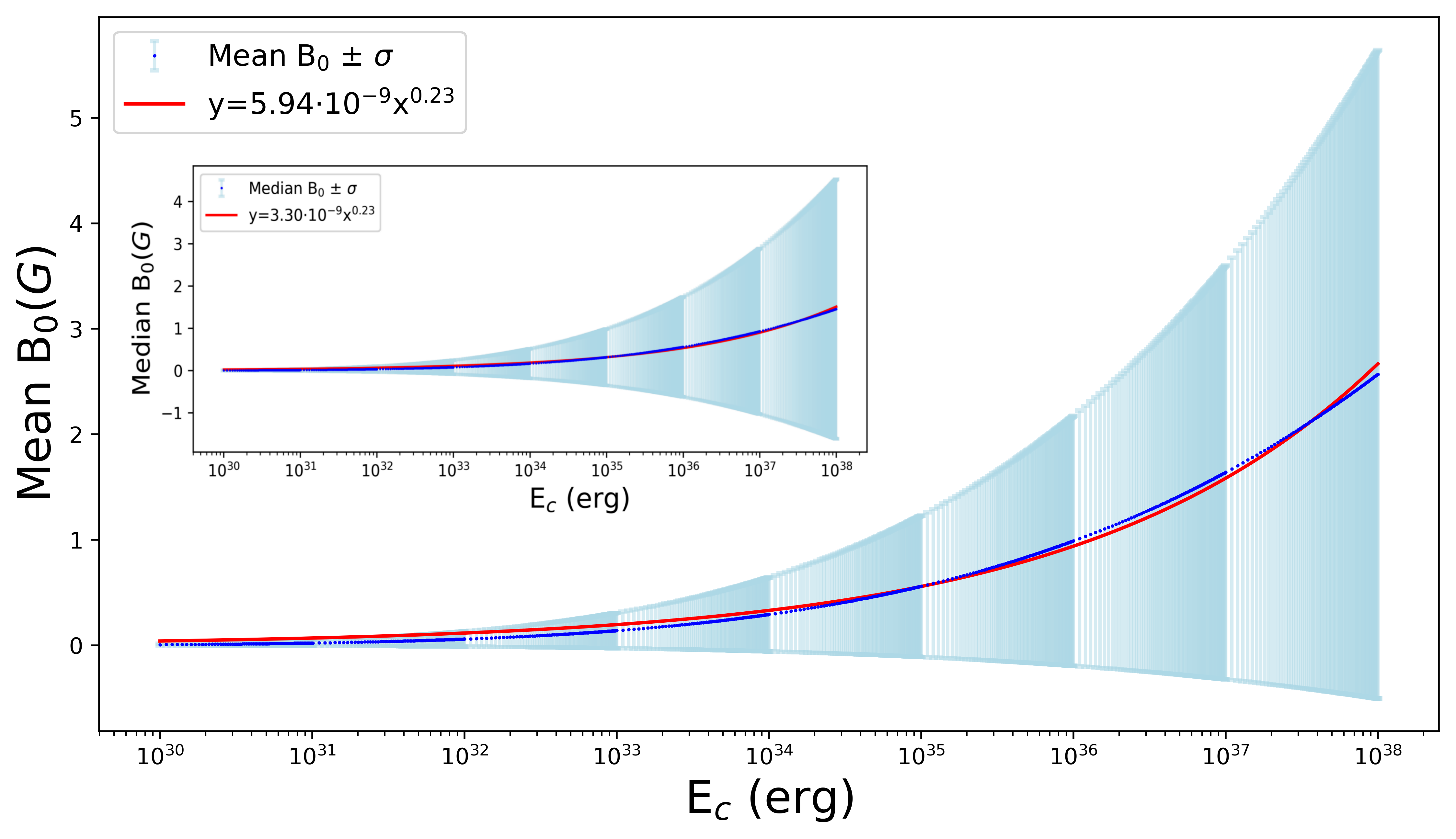}
    \caption{Mean (main plot) and median (inset) values of the near-star CME axial magnetic field $B_0$ at a physical distance of 10 $R_{\odot}$ from the star. The modeled data are shown by blue points, while least-squares power-law best fits are shown by red curves. Standard deviations around each $B_0$-value are represented by cyan segments.}
    \label{fig:nearSun_B0}
\end{figure}

Figure \ref{fig:nearSun_B0} provides the mean $B_{0_{(10\;R_{\odot})}}$ of the above-mentioned Monte Carlo simulation for different flare energies $E_c$, along with a standard deviation around these values (cyan ranges). The inset in Figure \ref{fig:nearSun_B0} provides the respective median values. We notice that both mean and median $B_{0_{(10\;R_{\odot})}}$ can be adequately modeled by power laws of the flare energy $E_c$ of the form 

\begin{equation}
B_{0_{(10\;R_{\odot})}}\;(G)\;=\;f\;E_c^{0.23}\;(erg)\;\;, 
\label{B0ns}
\end{equation}

\noindent where the proportionality constant $f$ is $3.30 \times 10^{-9}$ (median) or $5.94 \times 10^{-9}$ (mean), with an uncertainty amplitude 1.29 $\times 10^{-8}$.


\section{ICME equipartition magnetic field}
\label{appendix:ICMEequipartition}

\begin{figure}[ht]
    \centering
    \includegraphics[width=0.7\textwidth]{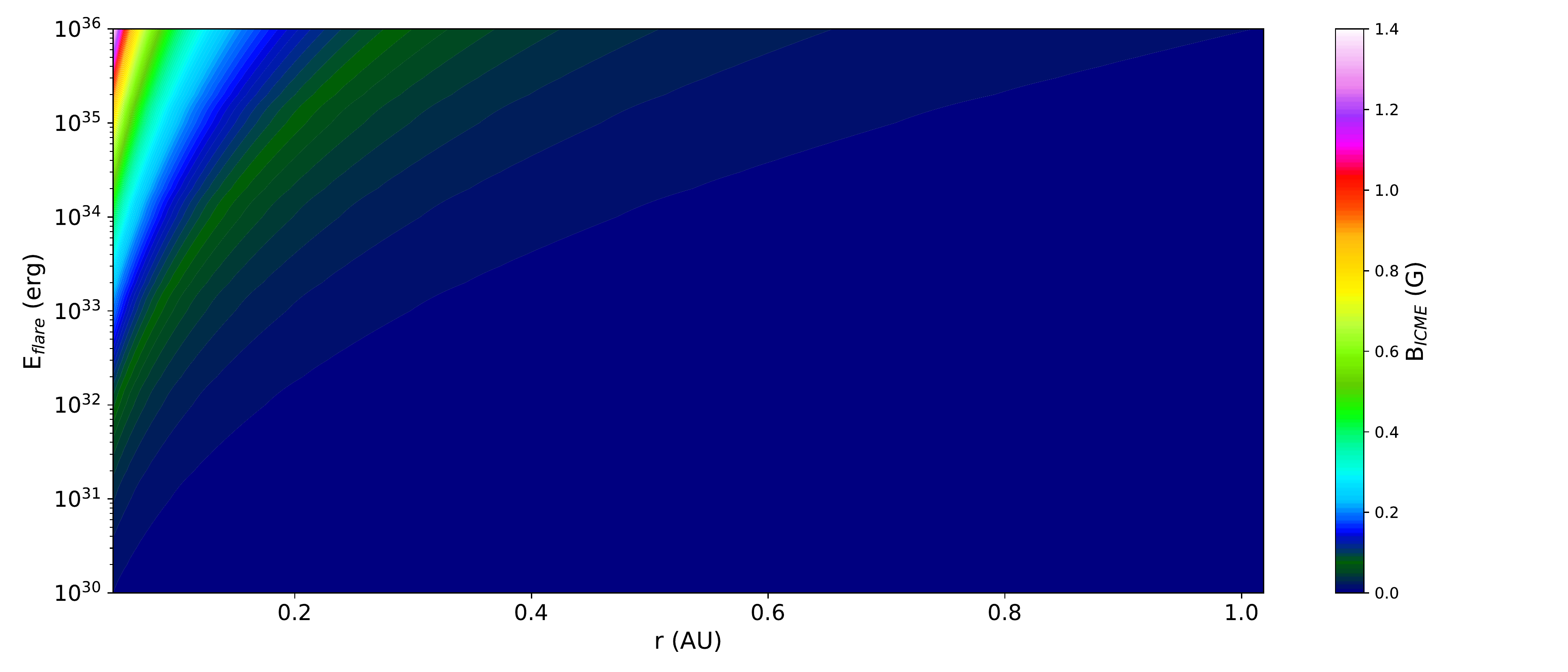}
    \caption{Worst-case scenario axial magnetic field for ICME magnetic flux ropes as a function of source flare energy and astrocentric distance, up to one astronomical unit (AU). A wide range of flare energies are provided, from flares observed in the Sun (i.e., up to $10^{33}$ erg) to orders of magnitude stronger superflares with energies up to $10^{36}$ erg.}
    \label{fig:bicme}
\end{figure}

Figure \ref{fig:bicme} provides the worst-case scenario axial magnetic field of ICME magnetic flux ropes (as virtually all CMEs/ICMEs are expected to be) as a function of solar/stellar flare energies and helio-/astro-centric distances. The ICME magnetic pressure effects on the planet will be determined from this magnetic field strength. 

\begin{figure}[ht]
    \centering
    \includegraphics[width=0.6\textwidth]{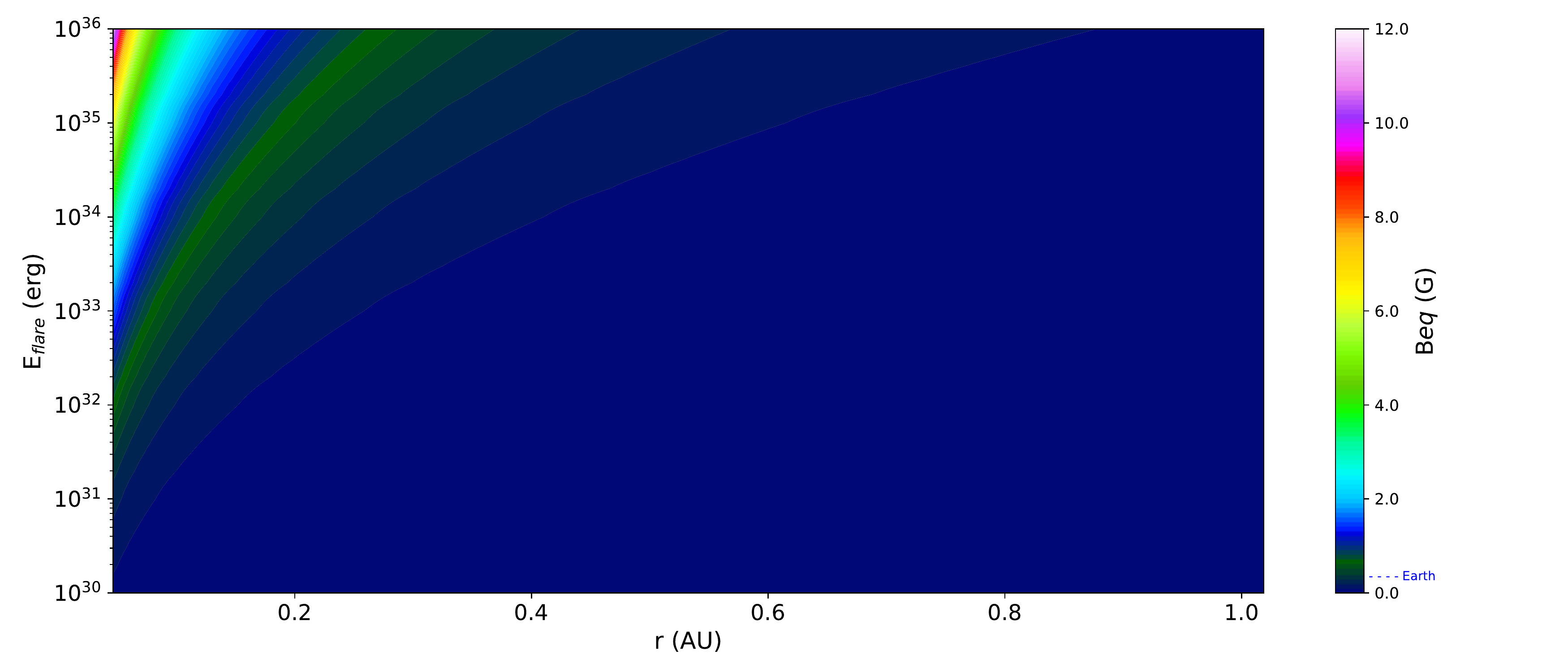}
    \caption{Same as in Figure \ref{fig:bicme} but showing the equipartition magnetic field of the planet at the adopted atmospheric-erosion critical threshold of two planetary radii. An indication labeled 'Earth' in the color bar shows the uncompressed terrestrial equatorial magnetic field, for reference.}
    \label{fig:beq}
\end{figure}

Textbook physics dictates that either nominal solar/stellar winds or 'stormy' ICMEs exert pressure that comprises magnetic (i.e., $B^2/(8 \pi)$), kinetic (i.e., ram; $\rho \upsilon ^2$) and thermal (i.e., $n\;k\;T$) terms as per the local magnetohydrodynamical (MHD) environment (plasma number density, mass density and speed, $n$, $\rho$ and $\upsilon$, respectively, and magnetic field $B$). At the planet's dayside at few to several planetary radii, the only non-negligible pressure term is magnetic pressure stemming from the planetary magnetosphere. This is because a possible atmosphere typically wanes at the planetary thermosphere or exosphere, already at small fractions of a  planetary radius. In a seminal early work, \citet{chapman1930} considered a spherical magnetosphere interfacing with interplanetary ejecta. Adopting this working hypothesis and taking only the magnetic pressure term of the ICME determined by its characteristic field $B_{icme}$, the existence of a magnetopause at planetocentric distance $r_{mp}$ implies an equilibrium of the form \citep{patgeo17}
\begin{equation}
{{B_{icme}^2} \over {8 \pi}} = {{B_{mp}^2} \over {8 \pi}} 
({{1} \over {r_{mp}}})^6\;\;,
\label{eq:mpeq}
\end{equation}
where $r_{mp}$ is expressed in planetary radii and $B_{mp}$ is the planetary magnetic field at the magnetopause. We now adopt a critical magnetopause distance, that is, the minimum planetocentric distance at extreme compression in which atmospheric ionization and erosion can still be averted at $r_{mp}=2\;R_{p}$ (i.e., two planetary radii, or one radius away from the surface of the planet). This is a limit already adopted by several previous studies \citep{khodachenko_etal07,lammer2007}. Therefore, the \emph{equipartition} planetary magnetic field $B_{eq}$ that balances the ICME magnetic pressure at $r_{mp}=2\;R_{p}$ is 
\begin{equation}
B_{eq} = 8\;B_{icme}\;\;,
\label{eq:eqf_appendix}
\end{equation}
from Equation (\ref{eq:mpeq}). This equipartition field for the same range of flare energies and astrocentric distances
(viewed in this case as the exoplanets' mean orbital distances in circular orbits) as in Figure \ref{fig:bicme} is provided in Figure \ref{fig:beq}. Equation (\ref{eq:eqf_appendix}) and Figure \ref{fig:beq}, therefore, provide the requirement for the planetary magnetic field such that planets avoid atmospheric erosion due to a given ICME magnetic pressure. 

\noindent It would worth mentioning at this point, that the toroidal component of the interplanetary magnetic field could become important in the case of fast rotators such as M-stars (e.g., \citet{Petit2008}, \citet{kay2016}, \citet{Villarreal2019}). In such case, an extra term is added in the pressure-balance equation at the sub-stellar point. However, we note that in the case of active M-dwarf stars the notion of a Parker spiral for the associated interplanetary magnetic field may not be fully applicable, given the much frequent CME activity than in the solar case, that could keep the spiral significantly perturbed virtually at all times.

\section{Best-case planetary magnetic field in tidally locked regimes}
\label{appendix:PlanetaryBinTidallyLockedExoplanets}

The planet's dipole magnetic moment $\mathcal{M}$ gives rise to a planetary magnetic field 

\begin{equation}
B_p = {{\mathcal{M}} \over {r_{mp}^3}}
\label{eq:bp}
\end{equation}

\noindent for the dayside magnetopause occurring at a planetocentric distance $r_{mp}$. The magnetic moment and resulting magnetic field in exoplanets is generally unknown and one may need to use a known planetary field benchmark (i.e., Earth's or another) to estimate the equilibrium magnetopause (or stand-off) distance between a planet and the stellar wind it encounters. For Earth, in particular, this distance is $\sim 10\;R_{\oplus}$ for the unperturbed solar wind, where $R_{\oplus}$ is Earth's radius. \citet{patgeo17} concluded that the terrestrial magnetopause cannot be compressed to a value smaller than $\sim 5\;R_{\oplus}$ by the magnetic pressure of ICMEs, an estimate which aligns with other findings of extreme magnetospheric compression \citep[e.g.,][]{russell2000}.

While our knowledge of exoplanet magnetic fields is virtually non-existent, for the subset of exoplanets that lie in the tidally-locked zone of their host stars \citep{grassmeier_etal04, khodachenko_etal07} we have an additional constraint: the 1:1 spin-orbit resonance (see Figure \ref{fig:chz} to visually locate all such exoplanets). It implies that tidally locked exoplanets have a self-rotating speed equal to the rotational speed around their host stars, in a synchronous rotation. While exceptions are possible, this study adopts the 1:1 spin-orbit resonance because it affords us an estimate of the angular self-rotation of the planet in case the orbital period around its host star is known from observations. Synchronous rotation statistically weakens the planetary magnetic dipole moment, but at the same time it enables its first-order estimation, further enabling an estimation of the planetary magnetic field that is crucial for this analysis. In case of planets that are not tidally locked to their mother star, or we do not employ tidal locking, the planetary magnetic field (B$_{best}$) will have to be hypothesized by assigning an ad hoc magnetic dipole moment in Equation \ref{eq:bp}. No scaling law is employed then, but we can use a benchmark planetary field, such as Earth's, for example.

Emphasizing on terrestrial planets, we calculate the upper-case $\mathcal{M}$ provided by \citet{stevenson_etal83} model by assuming a core conductivity $\sigma = 5 \times 10^5\;S/m$ as per \citet{stevenson03}. The mean core density $\rho_c$ that enters the relationship is further assumed equal to the mean density of the planet, i.e., $\rho _c = \rho = 3 M_p/ (4 \pi R_p^3)$ where $M_p$ and $R_p$ are the planetary mass and radius, respectively. The assumption allows a density estimate based on direct or implicit observational facts relevant to the terrestrial planet under study. Moreover, an angular rotation ($\omega$) equal to the planet's angular rotation around its host star is adopted, for presumed tidally locked planets. In case of planets that are not tidally locked, as explained initially, ad hoc planetary field benchmarks must be used.

Another obvious unknown is the planetary core radius, $R_c$. For Earth, we have $R_c \simeq 0.55\;R_{\oplus}$ while for Mercury the fraction is significantly larger ($R_c \simeq 0.85\;R_{\mercury}$). Regardless, an upper limit of $R_c$ is the radius of the studied planet. In the frame of adopting the best-case scenario for the planetary magnetic pressure, we will use this upper limit, adopting $R_c=R_p$. As an example, the above settings and Equation (\ref{eq:mmom}) for Earth imply an equatorial magnetic field $\sim 0.309\;G$, almost identical to the nominal terrestrial equatorial field of $\sim 0.305\;G$.

\section{Sensitivity analysis}
\label{appendix:SensitivityAnalysis}

As already explained, we are essentially interested in the ratio $\mathcal{R} = B_{eq}/B_{Stev}$ of the equipartition magnetic field $B_{eq}$ of a planet at magnetopause distance $r_{mp}=2\;R_{p}$ to the expected upper-limit, Stevenson magnetic field ($B_{Stev}$) for the planet. This paragraph details our approach to determine whether applicable uncertainties are capable of  changing the outcome of $\mathcal{R}>1$ or $\mathcal{R}<1$ for a given exoplanet. 

We start by asking what value of the radial fall-off power-law index $a_B$ is required for $\mathcal{R}=1$. Under this condition and Equation (\ref{eq:prop}), we find 
\begin{equation}
a_{B_{(\mathcal{R}=1)}} = {{\log (B_{stev}/(8\;B_0))} \over {\log (r_0/r_{icme})}}\;\;.
\label{eq:abcr}
\end{equation}
Evidently, if we find $\mathcal{R}>1$ for the nominal $a_B=1.6$, then $a_{B_{(\mathcal{R}=1)}} > a_B$, meaning that the ICME magnetic field must decay more abruptly than assumed to achieve $\mathcal{R}=1$ at $2\;R_p$. The opposite is the case if we find $\mathcal{R}<1$ for $a_B=1.6$ (i.e., $a_{B_{(\mathcal{R}=1)}} < a_B$).

Let us now assume an uncertainty $\delta B_0$ of the near-star magnetic field of the CME, $B_0$. This relates to the uncertainty $\delta a_B$ on $a_{B_{(\mathcal{R}=1)}}$ as follows: 

\begin{equation}
{{\delta a_B} \over {a_{B_{(\mathcal{R}=1)}}}} = {{1} \over {ln(8 B_0/B_{stev})}} {{\delta B_0} \over {B_0}}\;\;.
\label{eq:err_ab}
\end{equation}

\noindent From Equation (\ref{eq:hm0}) we can relate $\delta B_0$ to the uncertainty in the CME helicity, $\delta H_m$, as
\begin{equation}
{{\delta H_m} \over {\mid H_m \mid}} = 2 {{\delta B_0} \over {B_0}}\;\;,
\label{eq:err_hm1}
\end{equation}
\noindent and by using the EH diagram of Equation (\ref{eq:eh_2}) we can find another expression for $\delta H_m$, namely 
\begin{equation}
{{\delta H_m} \over {\mid H_m \mid}} = 
[ 
\beta ^2 ({{\delta E_c} \over {E_c}})^2 + 
(\ln{E_c})^2 \delta \beta ^2
]^{1/2} \simeq ln{E_c} \; \delta \beta\;\;, 
\label{eq:err_hm2}
\end{equation}

\noindent for $(\delta E_c/E_c) \lesssim 4$ and typical superflare energies $E_c \ge 10^{30}$ erg. In Equation (\ref{eq:err_hm2}) we have propagated the uncertainties for the free energy of the eruptive flare, $\delta E_c$, and the power-law index in the EH diagram of Equation (\ref{eq:eh_2}), $\delta \beta \simeq 0.05$. In Equation (\ref{eq:err_hm1}) we have further assumed that the forward-modeled GCS geometrical properties $R$ and $L$ of the CME carry no uncertainties, as even assuming $(\delta k /k)= (\delta w /w)=1$ the dominant error term is still $(\delta B_0 / B_0)$. As further shown in Equation (\ref{eq:err_hm2}), given the term depending on the logarithm of $E_c$ we may ignore the contribution of $\delta E_c$ for typical superflare energies $E_c \in (10^{30},10^{38})$ erg, thus simplifying the uncertainty equation. 

Combining Equations (\ref{eq:err_hm1}) and (\ref{eq:err_hm2}), we eliminate $(\delta H_m/\mid H_m \mid)$ and solve for $\delta B_0$ to find 

\begin{equation}
{{\delta B_0} \over {B_0}} \simeq {1 \over 2}\;ln{E_c}\;\delta \beta\;\;. 
\label{eq:db0}    
\end{equation}

\noindent Then, substituting Equation (\ref{eq:db0}) to Equation (\ref{eq:err_ab}) we reach the desired expression for $\delta a_B$ on $a_{B_{(R=1)}}$ as follows: 

\begin{equation}
{{\delta a_B} \over {a_{B_{(\mathcal{R}=1)}}}} \simeq {{ln E_c} \over 
{2 ln(8 B_0/B_{stev})}} \delta \beta \;\;.
\label{eq:dab}
\end{equation}

\noindent Equation (\ref{eq:dab}) allows us to assign an uncertainty to $a_{B_{(\mathcal{R}=1)}}$ and the relevant question is whether $a_{B_{(R=1)}} \ne 1.6$ beyond this uncertainty. In this case, our conclusion on $\mathcal{R}$ is unlikely to change. 
In addition, if $\mathcal{R}>1$ and $a_{B_{(\mathcal{R}=1)}} >2$, then  $a_{B_{(\mathcal{R}=1)}}$ may be deemed unrealistic because the ICME magnetic field is required to fall more abruptly in the astrosphere than the 'unobstructed' $a_B =2$ case for the near-star CME expansion in order to achieve $\mathcal{R}=1$. If $a_{B_{(\mathcal{R}=1)}} >2$ beyond applicable uncertainties, then our conclusion for $\mathcal{R}>1$ is considered solid, whereas in case $\mid a_{B_{(\mathcal{R}=1)}} - 2 \mid \le \delta a_B$ our conclusion is again unlikely to change but it is conceivable that $\mathcal{R} \le 1$ for a significantly steeper than 1.6 decrease of the ICME magnetic field in the astrosphere. 

\begin{figure}[t]
\centering
    \includegraphics[width=\textwidth]{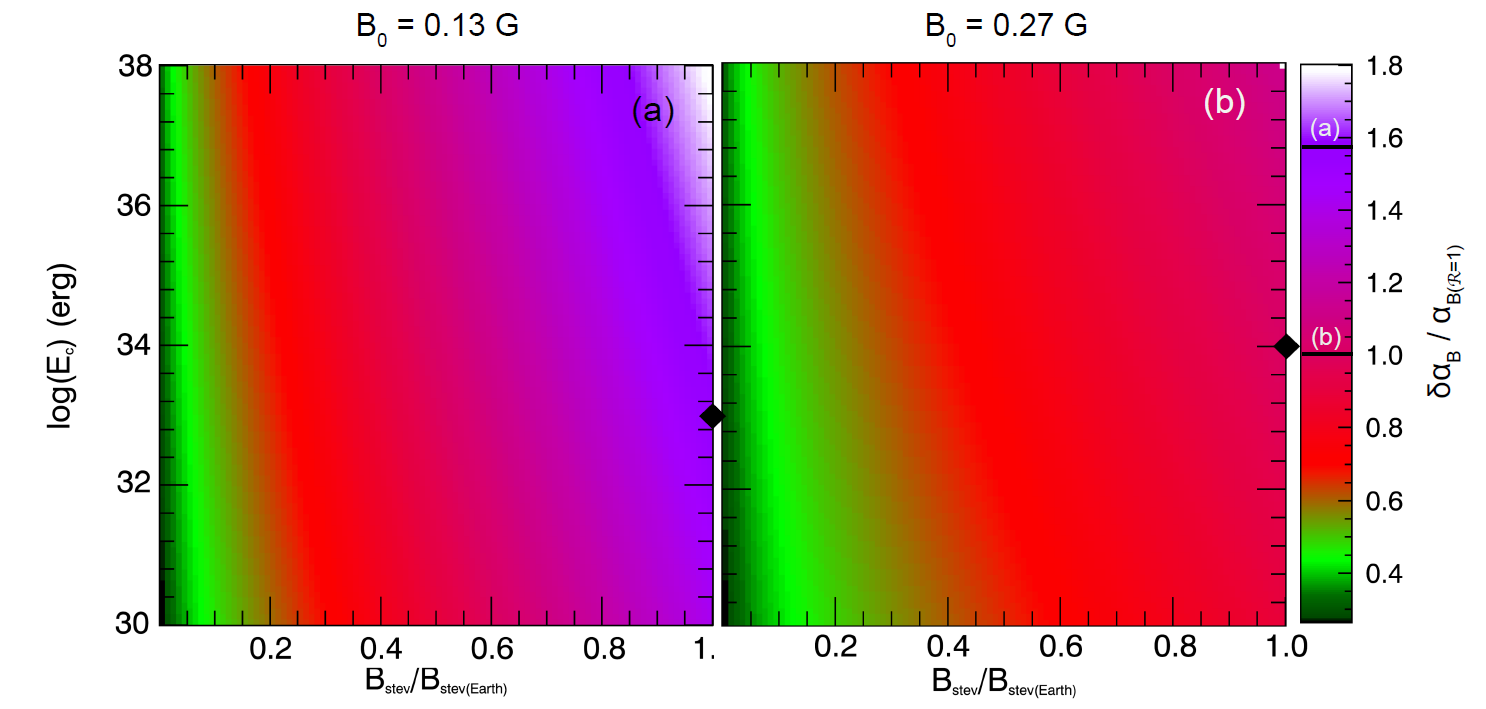}
    \caption{Normalized uncertainties $(\delta a_B / a_{B_{(\mathcal{R}=1)}})$ of the radial power-law fall-off index $a_B$ required for an equipartition ($\mathcal{R}=1$) between the ICME and planetary magnetic pressures at planetocentric distance equal to 2 planetary radii. $B_0$-values shown correspond to the mean near-Sun CME axial magnetic fields (Figure \ref{fig:nearSun_B0}) for $E_c =10^{33}$ erg (a) and $E_c = 10^{34}$ erg (b). Diamonds in images and indications in the color bar show $(\delta a_B / a_{B_{(\mathcal{R}=1)}})$ for Earth for  the two cases.}
    \label{fig:unc_ab}
\end{figure}
Figure \ref{fig:unc_ab} provides two examples of the ratio 
$(\delta a_B / a_{B_{(\mathcal{R}=1)}})$ 
of Equation (\ref{eq:dab}) 
for a wide range of superflare energies and Stevenson planetary fields ranging 
from 0 to the $B_{Stev}$-value for Earth ($B_{Stev_{\oplus}}\simeq 0.309$ G). Examples are based on two nominal $B_0$-values; one for $E_c=10^{33}$ erg (Figure \ref{fig:unc_ab}a) and another for $E_c=10^{34}$ erg (Figure \ref{fig:unc_ab}b). These are mean values stemming from the Monte Carlo simulations of Figure \ref{fig:nearSun_B0} for these two energies. Both are toward the upper end of the distribution for near-Sun CME magnetic fields as found in \citet{patgeo16}. Stronger $B_0$ increase the value of $a_{B_{(\mathcal{R}=1)}}$ (Equation (\ref{eq:abcr})) but correspondingly decrease its uncertainty fraction (Equation (\ref{eq:dab})). 

As an example, for the strong terrestrial magnetic field we find  
$a_{B_{(\mathcal{R}=1)}} = 0.40 \pm 0.63$ (Equations (\ref{eq:abcr}) and (\ref{eq:dab})) assuming $E_c=10^{33}$ erg and a nominal $r_0 = 10\;R_{\odot}$. The respective value for $E_c=10^{34}$ erg is $a_{B_{(\mathcal{R}=1)}} = 0.63 \pm 0.64$. These are both very flat $a_B$-values, 
flatter beyond uncertainties than the nominal $a_B=1.6$ that gives 
$\mathcal{R} \simeq 0.026$ and $\simeq 0.052$, respectively, under the same settings.

\bibliography{references}{}
\bibliographystyle{aasjournal}

\end{document}